\documentclass[%
reprint,
amsmath,amssymb,
aps,
]{revtex4-2}

\usepackage{graphicx}
\usepackage{dcolumn}
\usepackage{bm}

\usepackage{braket}
\usepackage{graphicx}
\usepackage{amsmath}
\usepackage{color}
\begin{document}

\title{Electromagnetically Induced Transparency Spectra of Ladder Four-Level System with Quantum Frequency Mixing}%

\author{Sheng-Xian Xiao}
\affiliation{Chongqing Key Laboratory for Strongly Coupled Physics, Chongqing University, Chongqing, 401331, China}

\affiliation{Center of Modern Physics, Institute for Smart City of Chongqing University in Liyang, Liyang 213300, China}
\author{Tao Wang}
\thanks{corresponding author: tauwaang@cqu.edu.cn}
\affiliation{Chongqing Key Laboratory for Strongly Coupled Physics, Chongqing University, Chongqing, 401331, China}

\affiliation{Center of Modern Physics, Institute for Smart City of Chongqing University in Liyang, Liyang 213300, China}
\begin{abstract}
In this paper, we generalized the quantum frequency mixing technology to a ladder-type four-level system and studied its effect on electromagnetically induced transparency spectra. We found a secondary splitting of Autler-Townes splitting in the probing field transmission spectra, which could be understood by the effective Hamiltonian derived with multi-mode Floquet theory. The Frequency mixing scheme developed here enables continuous tunablity of the resonant frequency between upper levels, which facilitates the broad band sensing of AC field.  Furthermore, by introducing an additional periodic driving, we realize an effective model that two distinct quantum interference effects coexist: interference among Floquet channels and loop interference arising from closed coherent pathways. Both interference effects could be read out from the transmission spectra independently. The changing of the distance between double splitting peaks represents the interference of Floquet channels, while their asymmetric linewidth broadening is linked with the total effective phase of the loop. This not only provides complementary readout for extracting the phase of AC field, but also establishes a new paradigm for coherent control in multi-level quantum systems.

\end{abstract}
\maketitle
\section{introduction\label{s1}}
Four-level systems stand as cornerstone platforms in atomic, molecular, and optical physics, offering unique flexibility for manipulating quantum coherence \cite{QC-1,QC-2} and light-matter interactions \cite{LMI-1,LMI-2,LMI-3}. The optical field-driven control of such systems has undergone extensive experimental and theoretical investigations across quantum sensing \cite{QS}, quantum metrology \cite{QM1,QM2}, and quantum computing \cite{QC}, laying the foundation for numerous cutting-edge technologies and functional device. These include optical switches \cite{OS1,OS2}, precise quantum state manipulation \cite{QSM1,QSM2}, implementation of phase gates \cite{PS1,PS2}, control of spontaneous emission \cite{SE-1,SE-2,SE-3,SE-4,SE-5,SE-6}, control of space-dependent four-wave mixing \cite{FWM}, frequency up-conversion amplification without inversion \cite{FA} and high-precision electric field sensing \cite{ES-1,ES-2}, etc. Yet, the exploration of novel coherent control paradigms remains an active frontier, motivating further investigations to expand the capabilities of four-level system-based platforms.

In parallel, quantum frequency mixing (QFM) \cite{25-PhysRevX.12.021061} , as an emerging quantum technology, utilizes the nonlinear effect of the periodically driven Floquet quantum system to mix the target signal field with the bias AC field, achieving the detection of signals of any frequency and breaking through the limitations of traditional quantum sensors. This technology has been successfully applied to wide-field magnetic images \cite{QFM-1}, searching of high-frequency variations of fundamental constants \cite{QFM-2}, and high-resolution and wide-frequency-range magnetic spectroscopy \cite{QFM-3}, etc. To date, however, implementations of QFM have largely been confined to two-level systems, most notably the nitrogen-vacancy (NV) center in diamond.  The extension of QFM to multi-level systems, and the exploration of its interplay with more complex coherent control protocols, remain largely unexplored. 

In this paper, we theoretically introduced QFM into a ladder four-level system  and studied its effect on transmission spectra of the probing field, which has the famous EIT\cite{EIT}(ectromagnetically induced transparency)-ATS(Autler-Townes splitting) structure. The quantum frequency mixing scheme is realized by applying  a resonant local oscillator (LO) field and two far-detuning fields to couple the two upper states.  We discover that the ATS peaks splits again in the spectra, forming a double-ATS structure, which could be explained by an effective model derived from multi-mode Floquet theory. This method effectively extends the system’s resonant response to far-detuning signal fields without the need for additional energy levels, enabling continued and broadband frequency sensing. Based on the above model, we further incorporated periodic driving on the LO field, forming a dual-Floquet-driven Hamiltonian in the dressed-state picture. The effective model of this double driving Hamiltonian reveals two distinct and tunable quantum interference mechanisms: (i) interference among different Floquet channels, and (ii) loop interference arising from closed coherent pathways. These are respectively manifested in the phase-dependent splitting distance and the asymmetric spectral broadening of the double-ATS peaks. Our work thus bridges QFM with multi-level coherent control, offering new avenues for quantum sensing and manipulation.

This paper is arranged as follows. In Sec. \ref{s2} we introduce the QFM scheme and its transmission spectra in the ladder Four-level system. In Sec. \ref{s3} we achieve dual-Floquet-driving in dressed-state picture, and finally in Sec. \ref{s4} we summarize and discuss the main findings of our work.

\begin{figure*}
	\includegraphics[width=1\linewidth]{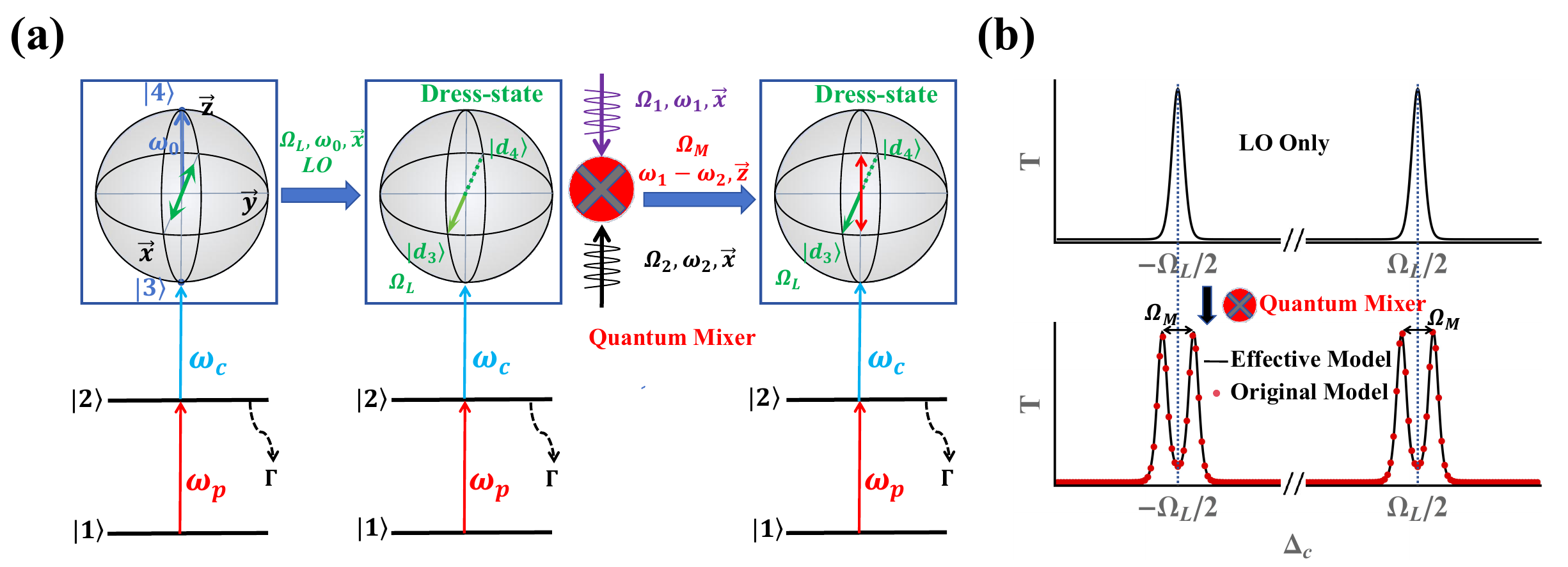}
	\caption{(a) Schematic diagram of quantum frequency mixing in the ladde four-level system, which consists of a ground state $\left| 1\right\rangle$, an excited state $\left| 2\right\rangle $ and two upper states $\left| 3\right\rangle $, $\left| 4\right\rangle$. The probe laser and the control laser are coupled to the transition $\left| 1\right\rangle \longleftrightarrow  \left| 2\right\rangle$ and $\left| 2\right\rangle \longleftrightarrow  \left| 3\right\rangle$. The two upper states are represented as the eigenstates in the $z$-axis of the Bloch sphere. The LO field along the $x$-axis induces the formation of the dressed states. Simultaneously, two far-detuning fields along the $x$-axis form an effective field along the $z$-axis through a quantum mixer, coupling the dressed states to form a new four-level system. (b) The transmission spectra when only the LO field exists (upper) and when an additional mixed-frequency field is resonant with two dressed states (lower). Parameters with the same unit: $\Omega_p/(2\pi)=\Omega_p/(2\pi)=2$, $\Gamma/(2\pi)=5$, $\Omega_L/(\pi)=80$, $\Delta_1/(2\pi)=1080$, $\Delta_2/(2\pi)=1000$, $\Omega_1/(2\pi)=108$, $\Omega_2/(2\pi)=10$.  \label{fig1}}
\end{figure*}

\section{The QFM scheme and transmission spectra \label{s2}}
We consider a ladder four-level system comprising a ground state $\left| 1\right\rangle$, an excited state $\left| 2\right\rangle $ and two metastable upper states $\left| 3\right\rangle $, $\left| 4\right\rangle $, as schematically depicted in Fig. \ref{fig1}(a). The excited state decays with rate $\Gamma$, while decay from the upper states is neglected. A probe laser with Rabi frequency $\Omega_p$ is resonant with transition $\left| 1\right\rangle \longleftrightarrow  \left| 2\right\rangle$. A control laser with Rabi frequency $\Omega_c$ and detuning $\Delta_c$ is coupled to the transition $\left| 2\right\rangle \longleftrightarrow  \left| 3\right\rangle$. Additionally, a strong resonant LO field with Rabi frequencie $\Omega_L$, and two far-detuning fields with Rabi frequencies $\Omega_1$, $\Omega_{2}$ and detuning $\Delta_1$, $\Delta_{2}$, are simultaneously coupled to the transition $\left| 3\right\rangle \longleftrightarrow  \left| 4\right\rangle$. 

Within the rotating-wave approximation (RWA), the system Hamiltonian is given by:
\begin{align}
    H=
    &-\frac{\hbar}{2}\left(\Omega_p\left| 1\right\rangle \left\langle 2\right|+\Omega_ce^{i\Delta_ct}\left| 2\right\rangle \left\langle 3\right|+h.c.\right)\nonumber\\
    &-\frac{\hbar}{2}\left[\left(\Omega_L+\Omega_{1}e^{i\Delta_{1}t}+\Omega_{2}e^{i\Delta_{2}t}\right)\left| 3\right\rangle \left\langle 4\right|+h.c.\right].\label{eq1}
\end{align}
The strong resonant LO field induces two dressed states  $\left| d_3\right\rangle $ and $\left| d_4\right\rangle$, which correspond to eigenstates along the $x$-axis in the Bloch sphere, with an energy gap of $\hbar \Omega_L$, as depicted in Fig.\ref{fig1} (a). The system could be seen as two effective three-level subsystem in dressed state picture: $\left| 1\right\rangle \longleftrightarrow  \left| 2\right\rangle \longleftrightarrow  \left| d_3\right\rangle$ and $\left| 1\right\rangle \longleftrightarrow  \left| 2\right\rangle \longleftrightarrow  \left| d_4\right\rangle$ by setting $\Delta_c$ to $\pm\Omega_L/2$, respectively. Consequently, when scanning  $\Delta_c$, the probe transmission spectrum exhibits two transparency windows, each featuring an ATS peak\cite{ES-1}. This constitutes the standard EIT-ATS spectrum in the presence of only the LO field, as shown in the upper panel of Fig.\ref{fig1} (b). 

The dressed states are related to the bare states by the transformation: $\left| d_3\right\rangle=\frac{1}{\sqrt{2}}\left(\left| 3\right\rangle-\left| 4\right\rangle\right)$ and $\left| d_4\right\rangle =\frac{1}{\sqrt{2}}\left(\left| 3\right\rangle+\left| 4\right\rangle\right)$. Substituting this into the original Hamiltonian Eq. (\ref{eq1}), and applying the QFM theory for the far-detuning fields , we derive an  effective Hamiltonian in the dressed-state picture (see Appendix \ref{A} for detailed calculation): 
\begin{align}
    H_d^{eff} =&-\frac{\hbar}{2}\Omega_p\left(\left| 1\right\rangle \left\langle 2\right|+h.c.\right)\nonumber\\
        &-\frac{\hbar}{2}\frac{\Omega_c}{\sqrt{2}}\left(e^{i\Delta_ct}\left| 2\right\rangle \left\langle d_3\right|+e^{i\Delta_ct}\left| 2\right\rangle \left\langle d_4\right|+h.c.\right)\nonumber\\
        &-\frac{\hbar}{2}\Omega_L\left(\left| d_4\right\rangle \left\langle d_4\right|-\left| d_3\right\rangle \left\langle d_3\right|\right)\nonumber\\        &+\hbar\left(\Omega_M\cos\omega_Mt+\delta_M\right)\left(\left| d_3\right\rangle \left\langle d_4\right|+h.c.\right),\label{eq2}
\end{align}
with effective mixed-frequency field parameters $\Omega_M=\Omega_1\Omega_2\left(\frac{1}{4\Delta_1}+\frac{1}{4\Delta_2}\right)$, $\omega_M=|\Delta_1-\Delta_2|$ and zero-frequency field parameter $\delta_M=\Omega_1^2/(4\Delta_1)+\Omega_2^2/(4\Delta_2)$ (also known as AC Stark shift). $\Omega_M$ can be linearly regulated by $\Omega_1$ and $\Omega_2$. Eq. (\ref{eq2}) reveals that  two far-detuning fields, through QFM, generate an effective field. In the Bloch sphere, this field is aligned along the $z$-axis with frequency $\omega$ and drives the transition between the dressed states, as illustrated in Fig.\ref{fig1} (a). 

Here we consider the case that the effective field resonates with the dressed states ($\omega_M=\Omega_L$). Applying a transformation $U=\exp[i\frac{\Omega_Lt}{2}(\left| d_4\right\rangle \left\langle d_4\right|-\left| d_3\right\rangle \left\langle d_3\right|)]$ to $H_d^{eff}$, for lager $\Omega_L\gg \Omega_M,2\delta_M$, we can obtain:
\begin{align}
    &U^\dagger\left(H_d^{eff}-i\hbar\frac{\partial}{\partial t}\right)U\overset{\text{ignoring $\Omega_L$ and $2\Omega_L$ terms }}{\Rightarrow }\nonumber\\
    =&-\frac{\hbar}{2}\Omega_p\left(\left| 1\right\rangle \left\langle 2\right|+h.c.\right)\nonumber\\
        &-\frac{\hbar}{2}\frac{\Omega_c}{\sqrt{2}}\left(e^{i\delta_c^-t}\left| 2\right\rangle \left\langle d_3\right|+e^{i\delta_c^+t}\left| 2\right\rangle \left\langle d_4\right|+h.c.\right)\nonumber\\
        &+\frac{\hbar}{2}\Omega_M\left(\left| d_3\right\rangle \left\langle d_4\right|+h.c.\right), \label{eqa9}
\end{align}
where $\delta_c^{\pm}=\Delta_c\pm\Omega_L/2$. We note that the zero-frequency term $\delta_M$ also corresponds a $z$-axis field but is far-detuning from the dressed-state transition due to the large $\Omega_L$ and is therefore neglected. Furthermore, the strong LO field ensures that the transitions  $\left| 2\right\rangle \longleftrightarrow  \left| d_3\right\rangle$ and $\left| 2\right\rangle \longleftrightarrow  \left| d_4\right\rangle$ cannot be simultaneously resonant for a given $\Delta_c$. Consequently, for $\Delta_c=\pm\Omega_L/2$, the system effectively reduces to two distinct four-level chains: $\left| 1\right\rangle \longleftrightarrow  \left| 2\right\rangle \longleftrightarrow  \left| d_3\right\rangle \longleftrightarrow  \left| d_4\right\rangle$ and $\left| 1\right\rangle \longleftrightarrow  \left| 2\right\rangle \longleftrightarrow  \left| d_4\right\rangle \longleftrightarrow  \left| d_3\right\rangle$. These two four-level chains can be analyzed independently and each of them supports a new ATS structure when the effective mixed-frequency field is resonant with the dressed-state splitting. The centers of these new ATS doublets remain at $\Delta_c=\pm\Omega_L/2$. This results in a secondary splitting of each original ATS peak, which we term the double-ATS. The splitting distance within this doublet is precisely given by the effective Rabi frequency $\Omega_M$. 

To obtain the transmission spectrum of our model, we simulate the system dynamic numerically using the master equation:
\begin{equation}
    \dot{\rho}=-\frac{i}{\hbar}[H,\rho]+\mathcal{L}(\rho),\label{eqb1}
\end{equation}
where $\mathcal{L}(\rho)$ is the Lindblad operator that accounts for the decay processes in the atom \cite{31-10.1063/1.4984201}. The matrix $\mathcal{L}(\rho)$ is given by:
\begin{align}
	\left[\begin{array}{cccc}
		\Gamma_{2} \rho_{22} & -\gamma_{12} \rho_{12} & -\gamma_{13} \rho_{13} & -\gamma_{14} \rho_{14} \\
		-\gamma_{21} \rho_{21} & \Gamma_{3} \rho_{33}-\Gamma_{2} \rho_{22} & -\gamma_{23} \rho_{23} & -\gamma_{24} \rho_{24} \\
		-\gamma_{31} \rho_{31} & -\gamma_{32} \rho_{32} & \Gamma_{4} \rho_{44}-\Gamma_{3} \rho_{33} & -\gamma_{34} \rho_{34} \\
		-\gamma_{41} \rho_{41} & -\gamma_{42} \rho_{42} & -\gamma_{43} \rho_{43} & -\Gamma_{4} \rho_{44}\label{eqb2}
	\end{array}\right],
\end{align}
where $\gamma_{ij}=(\Gamma_{i}+\Gamma_{j})/2$ and $\Gamma_{1,2,3,4}$ are the decay rate of the four levels. The transmission of the probe laser is determined by the density matrix component $\rho_{21}$ via: $T=\exp{\left[-\mathcal{K}\mathrm{Im}(\rho_{21})\right]}$, where $\mathcal{K}$ depends on parameters such as the probe laser intensity and wavelength, the atomic gas density, the length of the atomic vapor cell and the transition dipole moment between states $\left| 1\right\rangle$ and $\left| 2\right\rangle$. In this paper, we set $\mathcal{K}=50$ and don't take into account the Doppler effect for a conceptual demonstration. 

As shown in the lower panel of Fig. \ref{fig1} (b), we present the transmission spectra from numerical master equation simulations \cite{31-10.1063/1.4984201} of both the original model Eq. (\ref{eq1}) and effective model Eq. (\ref{eq2}), where the excellent agreement validates our theory.

\begin{figure}
	\includegraphics[width=0.99\linewidth]{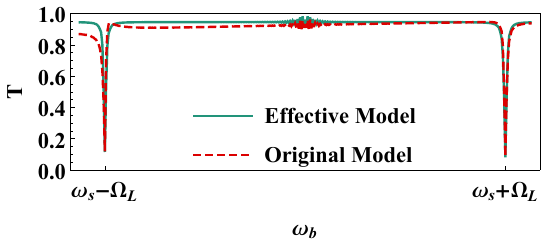}
	\caption{Transmission spectrum of scanning $\Delta_b$ for setting $\Delta_c=\Omega_L/2$. The scanning method is fixing $\Omega_b/\Delta_b=$0.1. $\Delta_s/(2\pi)=1000$,$\Omega_s/(2\pi)=10$, other parameters are same as in Fig. \ref{fig1} (b).\label{fig2}} 
\end{figure}

Crucially, the QFM effect endows the system with a resonant response to fields that are far-detuning from the bare atomic transitions. This enables the double-ATS structure to facilitate resonant sensing of signals across a wide, continuous frequency band, without requiring additional energy levels. To exploit this for sensing, we designate one far-detuning field as the signal (with detuning $\Delta_s$ ) and the other as a tunable bias (with detuning  $\Delta_b$). The LO field fixes the dressed-state splitting $\Omega_L$, Maximum response occurs when the beat note between the signal and bias fields matches this splitting, satisfying the resonance condition $\Delta_s-\Delta_b=\pm\Omega_L$. By fixing the control detuning at  $\Omega_L/2$ (or $-\Omega_L/2$) and scanning the bias frequency, the probe transmission spectrum will exhibit two resonant peaks,as shown in Fig.\ref{fig2}. Their positions, at $\Delta_b=\Delta_s-\Omega_L$ and $\Delta_b=\Delta_s+\Omega_L$, are directly determined by the unknown signal frequency $\Delta_s$. This dual-peak structure thus enables the simultaneous extraction of both the signal strength (from the peak amplitudes) and its frequency (from the peak positions), providing a complete protocol for detecting and characterizing signals of unknown frequency. 

\section{dual-Floquet-driving in dressed-state picture\label{s3}}
We now introduce a periodic modulation to the LO field, implementing a dual-Floquet drive in the dressed-state picture. This approach enables advanced control over the double-ATS and reveals new interference phenomena. The Hamiltonian incorporating a periodic driving of the frequency $\omega$, the strength $g$ and the phase $\phi$ on the LO field is given by:
\begin{align}
    H_d^{eff} =&-\frac{\hbar}{2}\Omega_p\left(\left| 1\right\rangle \left\langle 2\right|+h.c.\right)\nonumber\\
        &-\frac{\hbar}{2}\frac{\Omega_c}{\sqrt{2}}\left(e^{i\Delta_ct}\left| 2\right\rangle \left\langle d_3\right|+e^{i\Delta_ct}\left| 2\right\rangle \left\langle d_4\right|+h.c.\right)\nonumber\\
        &-\frac{\hbar\Omega_L}{2}\left[1+g\cos(\omega t+\phi)\right]\left(\left| d_4\right\rangle \left\langle d_4\right|-\left| d_3\right\rangle \left\langle d_3\right|\right)\nonumber\\
        &+\hbar\left[\Omega_M\cos(\omega_Mt+\phi_M)+\delta_M\right]\left(\left| d_3\right\rangle \left\langle d_4\right|+h.c.\right) \label{eq3}
\end{align}
Here, $\delta_M$ is included as its effect can become resonant via Floquet photon-assisted processes. Besides, the effective mixed-frequency phase $\phi_M$ are also considered.

\begin{figure}
	\includegraphics[width=0.99\linewidth]{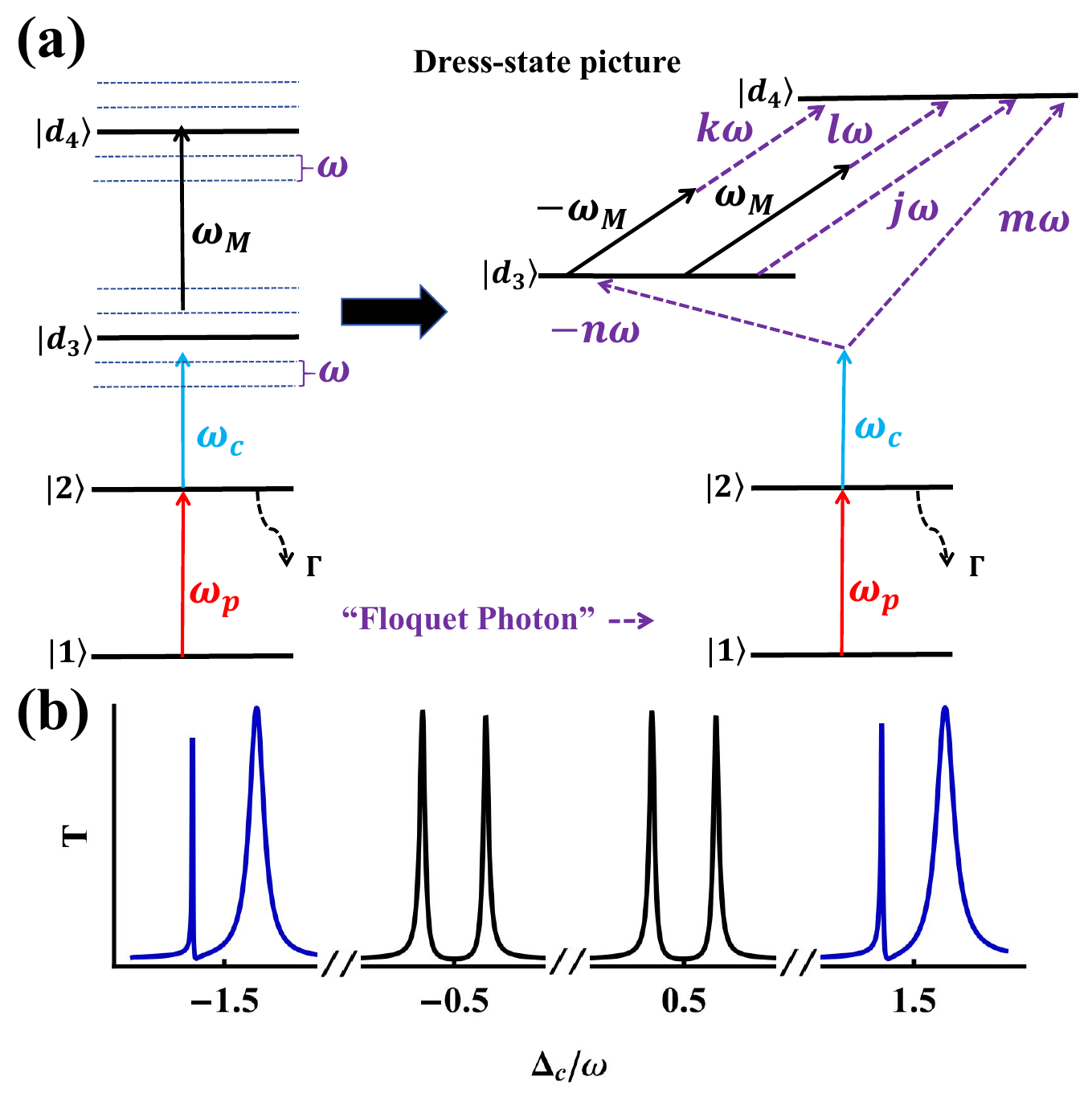}
	\caption{(a) Floquet photon-assisted transition of the four-level system in the dressed-state picture. (b) The transmission spectrum of double-ATS for different Floquet sideband. Parameters with the same unit: $\Omega_p/(2\pi)=0.1$, $\Omega_c/(2\pi)=5$, $\Gamma/(2\pi)=5$, $\Omega_L/(2\pi)=\omega/(2\pi)=40$, $\omega_M/(2\pi)=80$, $\Omega_M/(2\pi)=\delta_M/(2\pi)=4$. $g=4.81$, $\phi=\pi/2$, $\phi_M=0$. \label{fig3}} 
\end{figure}

The periodic modulation dresses the already-formed states $\left| d_3\right\rangle$ and $\left| d_4\right\rangle$ with an additional Floquet structure, creating sidebands separated by $\hbar\omega$ for each dressed state, as shown in the left of Fig. \ref{fig3} (a). Crucially, this allows far-detuned couplings to become resonant through the absorption or emission of Floquet photons, as illustrated for the Floquet photon-assisted transition in the right of Fig. \ref{fig3} (a). This periodic driving induces equal energy shifts in both dressed states--an advantage over Stark effect-driven bare states, where quadratic electric field dependence and differing polarizabilities cause unequal shifts, limiting controllability and resonant response.

The transmission spectrum under this dual driving, obtained by numerically solving the master equation for Eq. (\ref{eq3}), reveals a rich sideband structure. Fig. \ref{fig3} (b) presents the double-ATS spectra for different Floquet sideband. Notably, the spectra are not mere translations of the non-driven case. While the sideband at $\Delta_c=\pm0.5\omega$ remains symmetric, the sideband at $\Delta_c=\pm1.5\omega$ exhibits a striking asymmetry: one peak of the double-ATS broadens significantly while the other narrows.

This asymmetry stems from the formation of a closed coherent loop enabled by Floquet photons. When specific resonance conditions are met, the system dynamics are captured by a time-independent effective Hamiltonian (see Appendix \ref{C} for detailed calculation):
\begin{align}
    H_{n,m}^{l,k,j}=&-\frac{\hbar}{2}\Omega_p\left(\left| 1\right\rangle \left\langle 2\right|+h.c.\right)\nonumber\\
    &-\frac{\hbar}{2}\left(\Omega_c^n\left| 2\right\rangle \left\langle d_3\right|+\Omega_c^m\left| 2\right\rangle \left\langle d_4\right|+h.c.\right)\nonumber\\
    &+\frac{\hbar}{2}\left(|\Omega_M^{l,k,j}|e^{i\phi_M^{l,k,j}}\left| d_3\right\rangle \left\langle d_4\right|+h.c.\right) \label{eq4}
\end{align}
with Floquet photon number $n,m,l,k,j$, and effective parameters $\Omega_c^n=\frac{\Omega_c}{\sqrt{2}}J_n(\frac{g\Omega_L}{2\omega})$ and $\Omega_c^m=\frac{\Omega_c}{\sqrt{2}}J_m(\frac{g\Omega_L}{2\omega})$.  The  Floquet photon resonance conditions are $\Delta_c-\Omega_L/2-n\omega=0,\Delta_c+\Omega_L/2+m\omega=0,\omega_M+\Omega_L+l\omega=0$, $-\omega_M+\Omega_L+k\omega=0$ and $\Omega_L+j\omega=0$ . The complex coupling between dressed states, $|\Omega_M^{l,k,j}|e^{i\phi_M^{l,k,j}}$ ($|\Omega_M^{l,k,j}|>0$) synthesizes three distinct Floquet channels:
\begin{eqnarray}
\Omega_M^le^{i[\phi_M+(l-j)\phi)]}+\Omega_M^ke^{-i[\phi_M+(j-k)\phi]}+\delta_M^j,\label{eq5}
\end{eqnarray}
where $\Omega_M^l=\Omega_MJ_l(\frac{g\Omega_L}{\omega})$, $\Omega_M^k=\Omega_MJ_k(\frac{g\Omega_L}{\omega})$ and  $\delta_M^j=2\delta
_MJ_j(\frac{g\Omega_L}{\omega})$. Here, $J_n(x)$ represents the first kind of Bessel function with the order $n$. 

Eq. (\ref{eq4}) describes a system where states $\left| 2\right\rangle$, $\left| d_3\right\rangle$ and $\left| d_4\right\rangle$ are simultaneously coupled, forming a closed loop: $\left| 2\right\rangle \longleftrightarrow  \left| d_3\right\rangle \longleftrightarrow  \left| d_4\right\rangle \longleftrightarrow  \left| 2\right\rangle$. The interference within this loop, governed by the phase $\phi_{LI}=\phi_M^{l,k,j}$, directly controls the relative linewidths of the two double-ATS peaks. For a weak probe, the linewidths can be derived by solving the the steady-state solution of the maser equation (see Appendix \ref{D} for detailed calculation):
\begin{equation}
\begin{aligned}
     \Gamma_{left}=\frac{{\Omega_c^n}^2+{\Omega_c^m}^2}{4\Gamma}(1-\frac{2\Omega_c^n\Omega_c^m}{{\Omega_c^n}^2+{\Omega_c^m}^2}\cos\phi_{LI})\\
    \Gamma_{right}=\frac{{\Omega_c^n}^2+{\Omega_c^m}^2}{4\Gamma}(1+\frac{2\Omega_c^n\Omega_c^m}{{\Omega_c^n}^2+{\Omega_c^m}^2}\cos\phi_{LI})
\end{aligned}
\label{eq6}
\end{equation}
This explains the asymmetric broadening in Fig \ref{fig3} (b). The loop interference is maximized ($\phi_{LI}=0$ or $\pi$) when the two peaks are most dissimilar in linewidth and vanishes ($\phi_{LI}=\pi/2$ or $3\pi/2$) when they are symmetric. The loop interference can be completely suppressed by driving at parameters where $\Omega_c^n=0$ or $\Omega_c^n=0$ as is the case for the symmetric sideband at $\Delta_c=\pm0.5\omega$ in Fig. \ref{fig3} (b).

\begin{figure*}
	\includegraphics[width=0.32\linewidth]{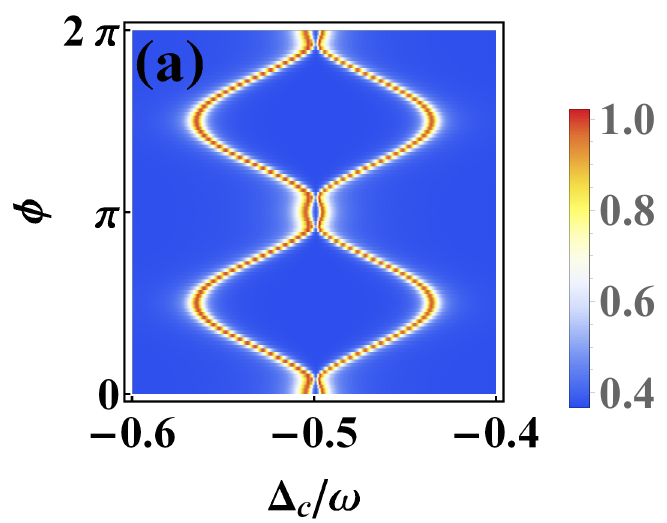}
    \includegraphics[width=0.32\linewidth]{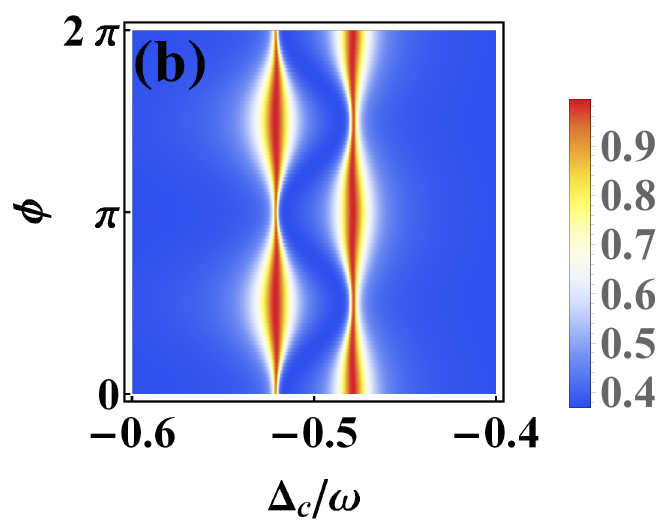}
    \includegraphics[width=0.32\linewidth]{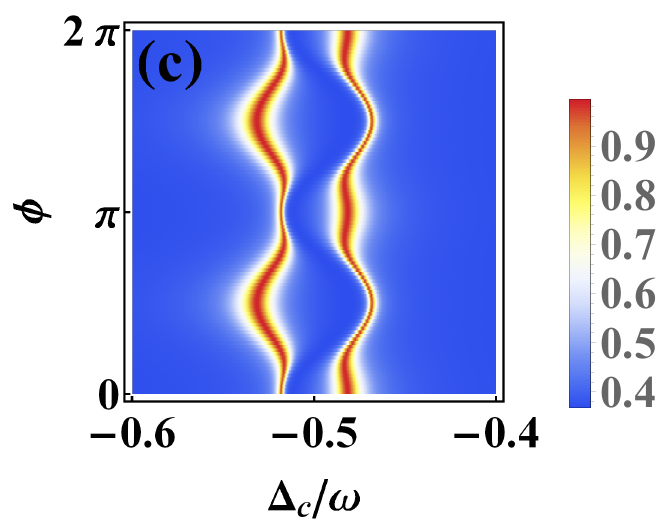}
	\caption{ The phase-tunable transmission spectra of double-ATS under three different conditions. (a) Floquet-channel interference only with $g=4.81$ ($\Omega_c^n\approx0$). (b) Loop interference only with $g=3.83$ ($\Omega_M^k,\delta_M^j\approx0$) . (c) Both Floquet-channel interference and Loop interference with $g=4$. Other parameters are the same as in Fig. \ref{fig3}. These parameters result in the Floquet photon number $n=-1, m=0,l=-3, k=1, j=-1$ for scanning center $\Delta_c=-0.5\omega$. \label{fig4}}
\end{figure*}
A second, independent interference effect arises among the three Floquet channels contributing to $|\Omega_M^{l,k,j}|$ in Eq.(\ref{eq5}): the zero-frequency term, the rotating wave term $\omega_M$ and the counter-rotating wave term $-\omega_M$,as shown in Fig. \ref{fig3} (a). The pairwise phase differences between these channels--$\phi_{FI}^{l,k}=2\phi_M+(l-k)\phi$, $\phi_{FI}^{j,l}=-\phi_M+(j-l)\phi$ and $\phi_{FI}^{k,j}=-\phi_M+(k-j)\phi$--determine the net coupling strength $|\Omega_M^{l,k,j}|$. This Floquet-channel interference manifests as a phase-dependent splitting distance between the double-ATS peaks:
\begin{equation}
\Delta f = \sqrt{
\begin{aligned}
\Omega_M^l{}^2 + \Omega_M^k{}^2 + \delta_M^j{}^2 
+ 2\Omega_M^l\Omega_M^k\cos\phi_{FI} ^{l,k}\\
+ 2\delta_M^j\Omega_M^l\cos\phi_{FI}^{j,l}+ 2\Omega_M^k\delta_M^j\cos\phi_{FI}^{k,j}
\end{aligned}
}\label{eq7}
\end{equation}
Similarly, the Floquet-channel interference will disappear when two of three parameters $\Omega_M^l$, $\Omega_M^k$ and $\delta_M^j$ are $0$ by adjusting the parameters of the driving field.


Notably, the phases of these two kinds of interference are not independent of each other. This is because the Floquet photon-assisted transition of the zero-frequency term links these two interference phases together, which can be seen from the Floquet photon number relationship: $m+n=j$ and $l-j=j-k$. These relationships result in the loop phase not including the phases of $m$ and $-n$ Floquet photons, but only depending on the phases of $l$, $k$, and $j$ Floquet photons, and result in Eq. (\ref{eq5}) being dependent only on the overall phase: $\phi_M+(l-j)\phi$. Both interferences are simultaneously dependent on this phase.

Although the phases of two interferences are not independent, their physical origins and spectral consequences are distinct. More importantly, their occurrence can be controlled independently by tailoring the drive strength $g$. In Fig. {\ref{fig4}}, we numerically calculate transmission spectra of the Hamiltonian Eq. (\ref{eq3}) under three different driving condition, demonstrating this independent controllability. Fig. {\ref{fig4}} (a) shows with $g=4.81$ chosen such that $\Omega_c^n\approx0$, loop interference is suppressed. The system behaves as a simple ladder four-level system, where the double-ATS peak spacing varies periodically with the drive phase $\phi$, while the linewidths remain symmetric. While in Fig. {\ref{fig4}} (b), $g=3.83$ such that $\Omega_M^k,\delta_M^j\approx0$, Floquet-channel interference is turned off. The double-ATS peak spacing remains constant, but the linewidths exhibit pronounced asymmetry that varies periodically with $\phi$. When $g=4$, both types of interference exist simultaneously. Both the peak spacing and the linewidth asymmetry of the double-ATS undergo complex periodic modulation with the drive phase $\phi$, showcasing the interplay of the two mechanisms, as shown in Fig. {\ref{fig4}} (c).

These results confirm that the dual-Floquet drive provides two independent "knobs"--one tuning the peak separation via Floquet-channel interference, and the other tuning the linewidth symmetry via loop interference--both addressable through the phase of an external periodic drive.

\section{conclusion and discussion\label{s4}}
In summary, we have successfully implemented quantum frequency mixing in a ladder four-level atomic system. This integration leads to a secondary splitting of the conventional ATS peaks, forming a distinct double-ATS structure in the EIT spectrum. The double-ATS enables resonant detection of signals across a broad, continuous frequency range without requiring additional atomic energy levels--effectively overcoming a fundamental bandwidth limitation in conventional quantum sensors. By further subjecting the system to a periodic Floquet drive, we have achieved dual-Floquet modulation and uncovered two independently tunable quantum interference mechanisms.

These interferences offer distinct capabilities for quantum control and sensing. Loop interference, arising from the artificially created closed coherent pathway:$\left| 2\right\rangle \longleftrightarrow  \left| d_3\right\rangle \longleftrightarrow  \left| d_4\right\rangle \longleftrightarrow  \left| 2\right\rangle$, provides a robust tool for controlling the lineshape of spectral features \cite{SE-5} and sensing the phase of microwave electric field \cite{LP1,LP2,LP3} without relying on the specific atomic natural  energy level transitions of atoms. The Floquet-channel interference, originating from the interplay between multiple Floquet channels, governs the effective coupling strength and thus the peak splitting distance. This allows for precise, phase-coherent tuning of the sensor's spectral resolution, facilitating tailored responses for specific signal analysis tasks. Crucially, both interference effects are phase-tunable via a common external driving field, yet their occurrence can be independently selected through the driving strength, establishing a versatile coherent control platform.

Our work bridges the fields of quantum frequency mixing and multi-level coherent control, expanding the toolkit for quantum state manipulation. The proposed scheme is particularly suited for implementation in Rydberg atoms or superconducting circuits, where strong dipole moments and precise microwave control are readily available. Future directions include exploring time-varying driving waveforms for dynamic sensing protocols and extending this framework to more complex quantum networks for quantum sensing.
\begin{acknowledgments}
This work is supported by National Science Foundation of China under Grant No. 12274045, No. 12274046 and No.12347101.
\end{acknowledgments}

\appendix

\section{effective Hamiltonian in the dressed-state picture\label{A}}
\subsection{QFM theory\label{A1}}
We consider a time-dependent Hamiltonian that can be described by two frequency modes $(\omega_a,\omega_b)$: 
\begin{eqnarray}
	H(t)=\sum_{m,n}H_{m,n}e^{im\omega_at}e^{im\omega_bt},\label{eqA1}
\end{eqnarray}
where $\omega_a$ and $\omega_b$ are much larger than other energy parameters. According to the theory of the quantum frequency mixing, the time-dependent effective Hamiltonian with low-frequency can be obtained as \cite{25-PhysRevX.12.021061}:
\begin{eqnarray}
	H^{eff}(t)=\sum_{l,k}\left( H_{l,k}+H^{(2)}_{l,k}+\dots\right)e^{i(l\omega_a+k\omega_b)t},\label{eqA2}
\end{eqnarray}
where summation indices $(l,k)$ satisfy the condition that $l\omega_a+k\omega_b\ll\omega_a,\omega_b$ and ${H^{(2)}_{l,k}}^\dagger=H^{(2)}_{-l,-k}$. The second-order term $H^{(2)}_{l,k}$ is given by:
\begin{eqnarray}
	-\frac{1}{2}\sum_{(p,q)\neq (l,k)}\frac{\left[ H_{l-p,k-q},H_{p,q}\right] }{p\hbar\omega_a+q\hbar\omega_b},\label{eqA3}
\end{eqnarray}
where the summation excludes the low-frequency case $(p,q)=(l,k)$. 
\subsection{Effective Hamiltonian\label{A2}}
The original system Hamiltonian Eq. (\ref{eq1}) is described by two frequency modes $(\Delta_1,\Delta_2)$ for large detunings $\Delta_1,\Delta_2\gg\Omega_p, \Omega_c,\Omega_L,\Delta_c, |\Delta_1-\Delta_2|$, and its Fourier components are:
\begin{equation}
    \begin{aligned}
        &H_{0,0}=-\frac{\hbar}{2}\left(\Omega_p\left| 1\right\rangle \left\langle 2\right|+\Omega_ce^{i\Delta_ct}\left| 2\right\rangle \left\langle 3\right|+\Omega_L\left| 3\right\rangle \left\langle 4\right|+h.c.\right)\\
        &H_{1,0}=-\frac{\hbar}{2}\Omega_1\left| 3\right\rangle \left\langle 4\right|, H_{-1,0}=-\frac{\hbar}{2}\Omega_1\left| 4\right\rangle \left\langle 3\right|\\
        &H_{0,1}=-\frac{\hbar}{2}\Omega_2\left| 3\right\rangle \left\langle 4\right|,H_{0,-1}=-\frac{\hbar}{2}\Omega_2\left| 4\right\rangle \left\langle 3\right|
    \end{aligned}
    \label{eqA4}
\end{equation}
for two frequency modes $(\Delta_1,\Delta_2)$, there are two kinds low-frequency mixing terms: the zero-frequency term and $\pm(\Delta_1-\Delta_2)$ term. Substituting Eq. (\ref{eqA4}) to Eq. (\ref{eqA3}), we can get the second-order term:
\begin{align}
    H_{1,-1}^{(2)}=&-\frac{1}{2}\left[\frac{[H_{1,0},H_{0,-1}]}{-\hbar\Delta_2}+\frac{[H_{0,-1},H_{1,0}]}{\hbar\Delta_1}\right]\nonumber\\
    &-\frac{\Omega_1\Omega_2}{8}\left(\frac{1}{\Delta_1}+\frac{1}{\Delta_2}\right)\left(\left| 4\right\rangle \left\langle 4\right|-\left| 3\right\rangle \left\langle 3\right|\right)
\end{align}
and
\begin{align}
    H_{0,0}^{(2)}=&-\frac{1}{2}\left[\frac{[H_{1,0},H_{-1,0}]}{-\hbar\Delta_1}+\frac{[H_{-1,0},H_{1,0}]}{\hbar\Delta_1}\right]\nonumber\\
    &-\frac{1}{2}\left[\frac{[H_{0,1},H_{0,-1}]}{-\hbar\Delta_2}+\frac{[H_{0,-1},H_{0,1}]}{\hbar\Delta_2}\right]\nonumber\\
    =&-(\frac{\Omega_1^2}{4\Delta_1}+\frac{\Omega_2^2}{4\Delta_2})\left(\left| 4\right\rangle \left\langle 4\right|-\left| 3\right\rangle \left\langle 3\right|\right)
\end{align}
Then, according to Eq. (\ref{eqA2}), the time-dependent effective Hamiltonian is given by:
\begin{align}
    H^{eff}=&H_{0,0}+H_{0,0}^{(2)}+\left[H_{1,-1}^{(2)}e^{i(\Delta_1-\Delta_2)t}+h.c.\right]\nonumber\\
    =&-\frac{\hbar}{2}\left(\Omega_p\left| 1\right\rangle \left\langle 2\right|+\Omega_ce^{i\Delta_ct}\left| 2\right\rangle \left\langle 3\right|+h.c.\right)\nonumber\\
    &-\frac{\hbar}{2}\left(\Omega_L\left| 3\right\rangle \left\langle 4\right|+h.c.\right)\nonumber\\
    &-\hbar\left(\Omega_M\cos\omega_Mt+\delta_M\right)\left(\left| 4\right\rangle \left\langle 4\right|-\left| 3\right\rangle \left\langle 3\right|\right)\label{eqa7},
\end{align}
where $\Omega_M=\Omega_1\Omega_2\left(\frac{1}{4\Delta_1}+\frac{1}{4\Delta_2}\right)$, $\omega_M=|\Delta_1-\Delta_2|$ and $\delta_M=\Omega_1^2/(4\Delta_1)+\Omega_2^2/(4\Delta_2)$.
Substituting the transformation: $\left| d_3\right\rangle=\frac{1}{\sqrt{2}}\left(\left| 3\right\rangle-\left| 4\right\rangle\right)$ and $\left| d_4\right\rangle =\frac{1}{\sqrt{2}}\left(\left| 3\right\rangle+\left| 4\right\rangle\right)$ into Eq. (\ref{eqa7}) we can get the effective Hamiltonian in the dressed-state picture:
\begin{align}
    H_d^{eff} =&-\frac{\hbar}{2}\Omega_p\left(\left| 1\right\rangle \left\langle 2\right|+h.c.\right)\nonumber\\
        &-\frac{\hbar}{2}\frac{\Omega_c}{\sqrt{2}}\left(e^{i\Delta_ct}\left| 2\right\rangle \left\langle d_3\right|+e^{i\Delta_ct}\left| 2\right\rangle \left\langle d_4\right|+h.c.\right)\nonumber\\
        &-\frac{\hbar}{2}\Omega_L\left(\left| d_4\right\rangle \left\langle d_4\right|-\left| d_3\right\rangle \left\langle d_3\right|\right)\nonumber\\        &+\hbar\left(\Omega_M\cos\omega_Mt+\delta_M\right)\left(\left| d_3\right\rangle \left\langle d_4\right|+h.c.\right) \label{eqa8}
\end{align}


\section{Derivation of Eq. (\ref{eq4})\label{C}}
Applying the following unitary transformation:
\begin{eqnarray}
    U_R=e^{i\left[\frac{\Omega_Lt}{2}+\frac{g\Omega_L}{2\omega}\sin(\omega t+\phi)\right]\left(\left| d_4\right\rangle \left\langle d_4\right|-\left| d_3\right\rangle \left\langle d_3\right|\right)}\label{eqc1}
\end{eqnarray}
into Eq. (\ref{eq3}), we can get:
\begin{widetext}
\begin{align}
    & U_R^{\dagger}\left(H_d^{eff}-i\hbar\frac{\partial}{\partial t}\right)U_R\nonumber\\
    =& -\frac{\hbar}{2}\Omega_p\left(\left| 1\right\rangle \left\langle 2\right|+h.c.\right)\nonumber\\
    & -\frac{\hbar}{2}\frac{\Omega_c}{\sqrt{2}}\left(e^{i\delta_c^-t}e^{-i\frac{g\Omega_L}{2\omega}\sin{(\omega t+\phi)}}\left| 2\right\rangle \left\langle d_3\right|
    +e^{i\delta_c^+t}e^{i\frac{g\Omega_L}{2\omega}\sin{(\omega t+\phi)}}\left| 2\right\rangle \left\langle d_4\right|+h.c.\right)\nonumber\\
    &+\hbar\Omega_M\cos(\omega_Mt+\phi_M)
    \left(e^{i\Omega_Lt}e^{i\frac{g\Omega_L}{\omega}\sin{(\omega t+\phi)}}\left| d_3\right\rangle \left\langle d_4\right|+h.c.\right)\nonumber\\
    &+\hbar\delta_Me^{i\Omega_Lt}e^{i\frac{g\Omega_L}{\omega}\sin{(\omega t+\phi)}}\left(\left| d_3\right\rangle \left\langle d_4\right|+h.c.\right)\label{eqc2}.
\end{align}
\end{widetext}
Substituting the identity $e^{iz\sin(\omega t)}=\sum_nJ_n(z)e^{in\omega t}$ into Eq. (\ref{eqc2}), and considering the Floquet photon resonance conditions $\Delta_c-\Omega_L/2-n\omega=0,\Delta_c+\Omega_L/2+m\omega=0,\omega_M+\Omega_L+l\omega=0$, $-\omega_M+\Omega_L+k\omega=0$ and $\Omega_L+j\omega=0$, we can get the effective Hamiltonian:
\begin{align}
    H_{n,m}^{l,k,j}=&-\frac{\hbar}{2}\Omega_p\left(\left| 1\right\rangle \left\langle 2\right|+h.c.\right)\nonumber\\
    &-\frac{\hbar}{2}\left(\Omega_c^ne^{-in\phi}\left| 2\right\rangle \left\langle d_3\right|+\Omega_c^me^{im\phi}\left| 2\right\rangle \left\langle d_4\right|+h.c.\right)\nonumber\\
    &+\frac{\hbar}{2}\left(|\Omega_M^{l,k,j}|e^{i(\phi_M^{l,k,j}+j\phi)}\left| d_3\right\rangle \left\langle d_4\right|+h.c.\right) \label{eqc3}
\end{align}
for $\omega\gg \Omega_c^{n^\prime} (n^\prime\neq n),\Omega_M^{l^{\prime}}  (l^\prime\neq l),\delta_M^{j^\prime} (j^\prime\neq j)$. Then applying the unitary transformation $U_\phi^\dagger H_{n,m}^{l,k,j}U_\phi$ with $U_\phi=e^{in\phi\left| d_3\right\rangle \left\langle d_3\right|-i m\phi\left| d_4\right\rangle \left\langle d_4\right|}$ and using the relation $j=m+n$, we can derive Eq. (\ref{eq4}) in the main text.

\section{The double-ATS linewidths \label{D}}
Since each Floque sideband spectrum is independent, we only need to calculate the double-ATS linewidths of one sidebands. When scanning $\Delta_c$ within in one Floquet sideband ($n,m$), $\Delta_c-\Omega_L/2-n\omega=\Delta_c+\Omega_L/2+m\omega$, which indicates that for the two transitions $\left| 2\right\rangle \longleftrightarrow \left| d_3\right\rangle$ and $\left| 2\right\rangle \longleftrightarrow \left| d_4\right\rangle$, the detunings of the photon-assisted transitions are the same. Thus, we can consider the Hamiltonian with the loop formulation:
\begin{align}
    H=&-\frac{\hbar}{2}\Omega_p\left(\left| 1\right\rangle \left\langle 2\right|+h.c.\right)\nonumber\\
    &-\frac{\hbar}{2}\left(\Omega_{c_1}e^{i\Delta t}\left| 2\right\rangle \left\langle 3\right|+\Omega_{c_2}e^{i\Delta t}\left| 2\right\rangle \left\langle 4\right|+h.c.\right)\nonumber\\
    &+\frac{\hbar}{2}\left(|\Omega|e^{i\phi}\left| 3\right\rangle \left\langle 4\right|+h.c.\right) \label{eqd1}
\end{align}
or
\begin{align}
    H=&-\frac{\hbar}{2}\Omega_p\left(\left| 1\right\rangle \left\langle 2\right|+h.c.\right)\nonumber\\
    &-\frac{\hbar}{2}\left(\Omega_{c_1}\left| 2\right\rangle \left\langle 3\right|+\Omega_{c_2}\left| 2\right\rangle \left\langle 4\right|+h.c.\right)\nonumber\\
    &-\hbar\Delta(\left| 3\right\rangle \left\langle 3\right|+\left| 4\right\rangle \left\langle 4\right|)\nonumber\\
    &+\frac{\hbar}{2}\left(|\Omega|e^{i\phi}\left| 3\right\rangle \left\langle 4\right|+h.c.\right) \label{eqd2}
\end{align}
Substituting Eq. (\ref{eqd2}) into the master equation \ref{eqb1}, we can obtain the steady-state solutions for the weak probe condition. The ground-state approximation is $\rho_{11}\approx1$, $\rho_{22}=\rho_{33}=\rho_{44}=0$ and $\rho_{23}=\rho_{24}=\rho_{34}=0$. The imaginary part of the density matrix component $\rho_{21}$ corresponding to the probe transition is obtained as: 
\begin{widetext}
\begin{align}
    \mathrm{Im}(\rho_{21})=\frac{\Omega_p\Gamma(2\Delta+|\Omega|)^2(2\Delta-|\Omega|)^2}{(\Omega_{c_1}^2+\Omega_{c_2}^2)^2\left(2\Delta+\frac{2\Omega_{c_1}\Omega_{c_2}}{\Omega_{c_1}^2+\Omega_{c_2}^2}|\Omega|\cos{\phi}\right)^2+\Gamma^2(2\Delta+|\Omega|)^2(2\Delta-|\Omega|)^2}\label{eqd3}
\end{align}
\end{widetext}
This Equation indicates that when $\Delta=\pm|\Omega|/2$, $\mathrm{Im}(\rho_{21})$, which corresponds to the double ATS peaks, suggesting that the loop phase does not affect the distance of the double ATS peaks.

To derive the double-ATS linewidths, we consider the small detuning near the ATS peak,i.e.,$\Delta=\pm|\Omega|/2+\epsilon$ ($\epsilon\ll|\Omega|$). Substituting $\Delta=-|\Omega|/2+\epsilon$ (corresponding to the left peak of the double ATS) into Eq. (\ref{eqd3}), we can obtain:
\begin{widetext}
\begin{align}
    \mathrm{Im}(\rho_{21})&=\frac{\Omega_p\Gamma(2\epsilon)^2(2\epsilon-2|\Omega|)^2}{(\Omega_{c_1}^2+\Omega_{c_2}^2)^2\left(2\epsilon-|\Omega|+\frac{2\Omega_{c_1}\Omega_{c_2}}{\Omega_{c_1}^2+\Omega_{c_2}^2}|\Omega|\cos{\phi}\right)^2+\Gamma^2(2\epsilon)^2(2\epsilon-2|\Omega|)^2}\nonumber\\
    &\approx\frac{4\Omega_p\Gamma\epsilon^2}{\frac{(\Omega_{c_1}^2+\Omega_{c_2}^2)^2}{4}\left(1-\frac{2\Omega_{c_1}\Omega_{c_2}}{\Omega_{c_1}^2+\Omega_{c_2}^2}\cos{\phi}\right)^2+4\Gamma^2\epsilon^2}
    \label{eqd4}
\end{align}
\end{widetext}
Compared with the Lorenz function: $\propto \frac{1}{\Gamma^2+\delta^2}$ ($\Gamma$ is the linewidth), Eq. (\ref{eqd4}) indicates the linewidth on the left side of the double ATS is
\begin{equation}
    \Gamma_{left}=\frac{\Omega_{c_1}^2+\Omega_{c_2}^2}{4}\left(1-\frac{2\Omega_{c_1}\Omega_{c_2}}{\Omega_{c_1}^2+\Omega_{c_2}^2}\cos{\phi}\right)
\end{equation}
In the same way, we can get the linewidth on the right side of the double ATS:
\begin{equation}
    \Gamma_{right}=\frac{\Omega_{c_1}^2+\Omega_{c_2}^2}{4}\left(1+\frac{2\Omega_{c_1}\Omega_{c_2}}{\Omega_{c_1}^2+\Omega_{c_2}^2}\cos{\phi}\right)
\end{equation}
\bibliographystyle{apsrev4-2}
\bibliography{ref}

@article{QC-1,
title = {Behavior of quantum coherence of Ξ-type four-level atom under bang–bang control},
journal = {Optics Communications},
volume = {281},
number = {18},
pages = {4793-4799},
year = {2008},
issn = {0030-4018},
doi = {https://doi.org/10.1016/j.optcom.2008.06.027},
url = {https://www.sciencedirect.com/science/article/pii/S0030401808005737},
author = {Yan-Hui Wang and Liang Hao and Xiang Zhou and Gui Lu Long}
}

@article{QC-2,
  title = {Certification and Quantification of Multilevel Quantum Coherence},
  author = {Ringbauer, Martin and Bromley, Thomas R. and Cianciaruso, Marco and Lami, Ludovico and Lau, W. Y. Sarah and Adesso, Gerardo and White, Andrew G. and Fedrizzi, Alessandro and Piani, Marco},
  journal = {Phys. Rev. X},
  volume = {8},
  issue = {4},
  pages = {041007},
  numpages = {12},
  year = {2018},
  month = {Oct},
  publisher = {American Physical Society},
  doi = {10.1103/PhysRevX.8.041007},
  url = {https://link.aps.org/doi/10.1103/PhysRevX.8.041007}
}

@article{LMI-1,
  title = {Strongly Interacting Photons in a Nonlinear Cavity},
  author = {Imamo\ifmmode \bar{g}\else \={g}\fi{}lu, A. and Schmidt, H. and Woods, G. and Deutsch, M.},
  journal = {Phys. Rev. Lett.},
  volume = {79},
  issue = {8},
  pages = {1467--1470},
  numpages = {0},
  year = {1997},
  month = {Aug},
  publisher = {American Physical Society},
  doi = {10.1103/PhysRevLett.79.1467},
  url = {https://link.aps.org/doi/10.1103/PhysRevLett.79.1467}
}

@article{LMI-2,
  title = {Polariton analysis of a four-level atom strongly coupled to a cavity mode},
  author = {Rebi\ifmmode \acute{c}\else \'{c}\fi{}, S. and Parkins, A. S. and Tan, S. M.},
  journal = {Phys. Rev. A},
  volume = {65},
  issue = {4},
  pages = {043806},
  numpages = {11},
  year = {2002},
  month = {Mar},
  publisher = {American Physical Society},
  doi = {10.1103/PhysRevA.65.043806},
  url = {https://link.aps.org/doi/10.1103/PhysRevA.65.043806}
}

@article{LMI-3,
  title = {Strongly correlated photons generated by coupling a three- or four-level system to a waveguide},
  author = {Zheng, Huaixiu and Gauthier, Daniel J. and Baranger, Harold U.},
  journal = {Phys. Rev. A},
  volume = {85},
  issue = {4},
  pages = {043832},
  numpages = {13},
  year = {2012},
  month = {Apr},
  publisher = {American Physical Society},
  doi = {10.1103/PhysRevA.85.043832},
  url = {https://link.aps.org/doi/10.1103/PhysRevA.85.043832}
}

@article{QS,
  title = {Quantum sensing},
  author = {Degen, C. L. and Reinhard, F. and Cappellaro, P.},
  journal = {Rev. Mod. Phys.},
  volume = {89},
  issue = {3},
  pages = {035002},
  numpages = {39},
  year = {2017},
  month = {Jul},
  publisher = {American Physical Society},
  doi = {10.1103/RevModPhys.89.035002},
  url = {https://link.aps.org/doi/10.1103/RevModPhys.89.035002}
}

@article{QM1,
  title = {Quantum Metrology},
  author = {Giovannetti, Vittorio and Lloyd, Seth and Maccone, Lorenzo},
  journal = {Phys. Rev. Lett.},
  volume = {96},
  issue = {1},
  pages = {010401},
  numpages = {4},
  year = {2006},
  month = {Jan},
  publisher = {American Physical Society},
  doi = {10.1103/PhysRevLett.96.010401},
  url = {https://link.aps.org/doi/10.1103/PhysRevLett.96.010401}
}

@article{QM2,
  title={Advances in quantum metrology},
  author={Giovannetti, Vittorio and Lloyd, Seth and Maccone, Lorenzo},
  journal={Nature photonics},
  volume={5},
  number={4},
  pages={222--229},
  year={2011},
  publisher={Nature Publishing Group UK London},
  doi = {110.1038/nphoton.2011.35},
  url = {https://doi.org/10.1038/nphoton.2011.35}
}

@ARTICLE{QC,
title={Qudits and high-dimensional quantum computing},
  author={Wang, Yuchen and Hu, Zixuan and Sanders, Barry C and Kais, Sabre},
  journal={Frontiers in Physics},
  volume={8},
  pages={589504},
  year={2020},
  publisher={Frontiers Media SA},
  URL={https://www.frontiersin.org/journals/physics/articles/10.3389/fphy.2020.589504},
  DOI={10.3389/fphy.2020.589504},
  ISSN={2296-424X}
}

@article{OS1,
  title = {Coherence Switching in a Four-Level System: Quantum Switching},
  author = {Ham, Byoung S. and Hemmer, Philip R.},
  journal = {Phys. Rev. Lett.},
  volume = {84},
  issue = {18},
  pages = {4080--4083},
  numpages = {0},
  year = {2000},
  month = {May},
  publisher = {American Physical Society},
  doi = {10.1103/PhysRevLett.84.4080},
  url = {https://link.aps.org/doi/10.1103/PhysRevLett.84.4080}
}

@article{OS2,
  title = {Optical switching and bistability in four-level atomic systems},
  author = {Kumar, Pardeep and Dasgupta, Shubhrangshu},
  journal = {Phys. Rev. A},
  volume = {94},
  issue = {2},
  pages = {023851},
  numpages = {8},
  year = {2016},
  month = {Aug},
  publisher = {American Physical Society},
  doi = {10.1103/PhysRevA.94.023851},
  url = {https://link.aps.org/doi/10.1103/PhysRevA.94.023851}
}

@article{QSM1,
  title = {Hamiltonian design to prepare arbitrary states of four-level systems},
  author = {Li, Yi-Chao and Mart\'{\i}nez-Cerc\'os, D. and Mart\'{\i}nez-Garaot, S. and Chen, Xi and Muga, J. G.},
  journal = {Phys. Rev. A},
  volume = {97},
  issue = {1},
  pages = {013830},
  numpages = {10},
  year = {2018},
  month = {Jan},
  publisher = {American Physical Society},
  doi = {10.1103/PhysRevA.97.013830},
  url = {https://link.aps.org/doi/10.1103/PhysRevA.97.013830}
}

@article{QSM2,
  title = {Universal composite pulses for robust quantum state engineering in four-level systems},
  author = {Shi, Zhi-Cheng and Wang, Jian-Hui and Zhang, Cheng and Song, Jie and Xia, Yan},
  journal = {Phys. Rev. A},
  volume = {109},
  issue = {2},
  pages = {022441},
  numpages = {15},
  year = {2024},
  month = {Feb},
  publisher = {American Physical Society},
  doi = {10.1103/PhysRevA.109.022441},
  url = {https://link.aps.org/doi/10.1103/PhysRevA.109.022441}
}

@article{PS1,
  title = {Phase gate with a four-level inverted-Y system},
  author = {Joshi, Amitabh and Xiao, Min},
  journal = {Phys. Rev. A},
  volume = {72},
  issue = {6},
  pages = {062319},
  numpages = {5},
  year = {2005},
  month = {Dec},
  publisher = {American Physical Society},
  doi = {10.1103/PhysRevA.72.062319},
  url = {https://link.aps.org/doi/10.1103/PhysRevA.72.062319}
}

@article{PS2,
  title = {Highly entangled photons and rapidly responding polarization qubit phase gates in a room-temperature active Raman gain medium},
  author = {Hang, Chao and Huang, Guoxiang},
  journal = {Phys. Rev. A},
  volume = {82},
  issue = {5},
  pages = {053818},
  numpages = {7},
  year = {2010},
  month = {Nov},
  publisher = {American Physical Society},
  doi = {10.1103/PhysRevA.82.053818},
  url = {https://link.aps.org/doi/10.1103/PhysRevA.82.053818}
}

@article{SE-1,
  title = {Spectral Line Elimination and Spontaneous Emission Cancellation via Quantum Interference},
  author = {Zhu, Shi-Yao and Scully, Marlan O.},
  journal = {Phys. Rev. Lett.},
  volume = {76},
  issue = {3},
  pages = {388--391},
  numpages = {0},
  year = {1996},
  month = {Jan},
  publisher = {American Physical Society},
  doi = {10.1103/PhysRevLett.76.388},
  url = {https://link.aps.org/doi/10.1103/PhysRevLett.76.388}
}

@article{SE-2,
  title = {Phase Control of Spontaneous Emission},
  author = {Paspalakis, E. and Knight, P. L.},
  journal = {Phys. Rev. Lett.},
  volume = {81},
  issue = {2},
  pages = {293--296},
  numpages = {0},
  year = {1998},
  month = {Jul},
  publisher = {American Physical Society},
  doi = {10.1103/PhysRevLett.81.293},
  url = {https://link.aps.org/doi/10.1103/PhysRevLett.81.293}
}

@article{SE-3,
  title = {Amplitude and phase control of spontaneous emission},
  author = {Ghafoor, Fazal and Zhu, Shi-Yao and Zubairy, M. Suhail},
  journal = {Phys. Rev. A},
  volume = {62},
  issue = {1},
  pages = {013811},
  numpages = {7},
  year = {2000},
  month = {Jun},
  publisher = {American Physical Society},
  doi = {10.1103/PhysRevA.62.013811},
  url = {https://link.aps.org/doi/10.1103/PhysRevA.62.013811}
}

@article{SE-4,
  title = {Control of spontaneous emission from a coherently driven four-level atom},
  author = {Wu, Jin-Hui and Li, Ai-Jun and Ding, Yue and Zhao, Yan-Chun and Gao, Jin-Yue},
  journal = {Phys. Rev. A},
  volume = {72},
  issue = {2},
  pages = {023802},
  numpages = {5},
  year = {2005},
  month = {Aug},
  publisher = {American Physical Society},
  doi = {10.1103/PhysRevA.72.023802},
  url = {https://link.aps.org/doi/10.1103/PhysRevA.72.023802}
}

@article{SE-5,
  title = {Effects of spontaneously generated coherence in a microwave-driven four-level atomic system},
  author = {Li, Ai-Jun and Song, Xiao-Li and Wei, Xiao-Gang and Wang, Lei and Gao, Jin-Yue},
  journal = {Phys. Rev. A},
  volume = {77},
  issue = {5},
  pages = {053806},
  numpages = {8},
  year = {2008},
  month = {May},
  publisher = {American Physical Society},
  doi = {10.1103/PhysRevA.77.053806},
  url = {https://link.aps.org/doi/10.1103/PhysRevA.77.053806}
}

@article{SE-6,
  title = {Spatiospectral control of spontaneous emission},
  author = {Asadpour, Seyyed Hossein and Abbas, Muqaddar and Hamedi, Hamid R. and Ruseckas, Julius and Paspalakis, Emmanuel and Asgari, Reza},
  journal = {Phys. Rev. A},
  volume = {110},
  issue = {3},
  pages = {033706},
  numpages = {11},
  year = {2024},
  month = {Sep},
  publisher = {American Physical Society},
  doi = {10.1103/PhysRevA.110.033706},
  url = {https://link.aps.org/doi/10.1103/PhysRevA.110.033706}
}

@article{FWM,
  title = {Control of space-dependent four-wave mixing in a four-level atomic system},
  author = {Qiu, Jing and Wang, Zhiping and Ding, Dongsheng and Huang, Zhixiang and Yu, Benli},
  journal = {Phys. Rev. A},
  volume = {102},
  issue = {3},
  pages = {033516},
  numpages = {6},
  year = {2020},
  month = {Sep},
  publisher = {American Physical Society},
  doi = {10.1103/PhysRevA.102.033516},
  url = {https://link.aps.org/doi/10.1103/PhysRevA.102.033516}
}

@article{FA,
  title = {Quantum-jump analysis of frequency up-conversion amplification without inversion in a four-level scheme},
  author = {Rubio, J. L. and Mompart, J. and Ahufinger, V.},
  journal = {Phys. Rev. A},
  volume = {107},
  issue = {1},
  pages = {013707},
  numpages = {9},
  year = {2023},
  month = {Jan},
  publisher = {American Physical Society},
  doi = {10.1103/PhysRevA.107.013707},
  url = {https://link.aps.org/doi/10.1103/PhysRevA.107.013707}
}

@article{ES-1,
	author = {Sedlacek, Jonathon A. and Schwettmann, Arne and Kübler, Harald and Löw, Robert and Pfau, Tilman and Shaffer, James P.},
	title = {Microwave electrometry with Rydberg atoms in a vapour cell using bright atomic resonances},
	journal = {Nature Physics},
	volume = {8},
	pages = {819–824},
	year = {2012},
	doi = {10.1038/nphys2423},
	URL = {https://doi.org/10.1038/nphys2423}
}

@article{ES-2,
	author = {Jing, Mingyong and Hu, Ying and Ma, Jie and Zhang, Hao and Zhang, Linjie and Xiao, Liantuan and Jia, Suotang },
	title = {Atomic superheterodyne receiver based on microwave-dressed Rydberg spectroscopy},
	journal = {Nature Physics},
	volume = {16},
	pages = {911–915},
	year = {2020},
	doi = {10.1038/s41567-020-0918-5},
	URL = {https://doi.org/10.1038/s41567-020-0918-5}
}

@article{25-PhysRevX.12.021061,
  title = {Sensing of Arbitrary-Frequency Fields Using a Quantum Mixer},
  author = {Wang, Guoqing and Liu, Yi-Xiang and Schloss, Jennifer M. and Alsid, Scott T. and Braje, Danielle A. and Cappellaro, Paola},
  journal = {Phys. Rev. X},
  volume = {12},
  issue = {2},
  pages = {021061},
  numpages = {22},
  year = {2022},
  month = {Jun},
  publisher = {American Physical Society},
  doi = {10.1103/PhysRevX.12.021061},
  url = {https://link.aps.org/doi/10.1103/PhysRevX.12.021061}
}

@article{QFM-1,
  title = {Quantum frequency mixing using an $\mathrm{N}$-$V$ diamond microscope},
  author = {Karlson, Samuel J. and Kehayias, Pauli and Schloss, Jennifer M. and Maccabe, Andrew C. and Libson, Adam and Phillips, David F. and Wang, Guoqing and Cappellaro, Paola and Braje, Danielle A.},
  journal = {Phys. Rev. Appl.},
  volume = {22},
  issue = {6},
  pages = {064051},
  numpages = {9},
  year = {2024},
  month = {Dec},
  publisher = {American Physical Society},
  doi = {10.1103/PhysRevApplied.22.064051},
  url = {https://link.aps.org/doi/10.1103/PhysRevApplied.22.064051}
}

@article{QFM-2,
    author = {Kong, Xi and Zhang, Yuke and Ji, Chenyu and Chang, Shuangju and Chen, Yifan and Bian, Xiang and Duan, Chang-Kui and Huang, Pu and Du, Jiangfeng},
    title = {Search of high-frequency variations of fundamental constants using spin-based quantum sensors},
    journal = {National Science Review},
    volume = {12},
    number = {4},
    pages = {nwaf085},
    year = {2025},
    month = {03},
    issn = {2095-5138},
    doi = {10.1093/nsr/nwaf085},
    url = {https://doi.org/10.1093/nsr/nwaf085},
}

@misc{QFM-3,
      title={High-resolution, Wide-frequency-range Magnetic Spectroscopy with Solid-state Spin Ensembles}, 
      author={Zechuan Yin and Justin J. Welter and Connor A. Hart and Paul V. Petruzzi and Ronald L. Walsworth},
      year={2024},
      eprint={2412.02040},
      archivePrefix={arXiv},
      primaryClass={quant-ph},
      url={https://arxiv.org/abs/2412.02040}, 
}

@article{EIT,
  title = {Electromagnetically induced transparency: Optics in coherent media},
  author = {Fleischhauer, Michael and Imamoglu, Atac and Marangos, Jonathan P.},
  journal = {Rev. Mod. Phys.},
  volume = {77},
  issue = {2},
  pages = {633--673},
  numpages = {0},
  year = {2005},
  month = {Jul},
  publisher = {American Physical Society},
  doi = {10.1103/RevModPhys.77.633},
  url = {https://link.aps.org/doi/10.1103/RevModPhys.77.633}
}

@article{31-10.1063/1.4984201,
    author = {Holloway, Christopher L. and Simons, Matt T. and Gordon, Joshua A. and Dienstfrey, Andrew and Anderson, David A. and Raithel, Georg},
    title = {Electric field metrology for SI traceability: Systematic measurement uncertainties in electromagnetically induced transparency in atomic vapor},
    journal = {Journal of Applied Physics},
    volume = {121},
    number = {23},
    pages = {233106},
    year = {2017},
    month = {06},
    issn = {0021-8979},
    doi = {10.1063/1.4984201},
    url = {https://doi.org/10.1063/1.4984201}
}

@article{LP1,
  title = {Optical Radio-Frequency Phase Measurement With an Internal-State Rydberg Atom Interferometer},
  author = {Anderson, D.A. and Sapiro, R.E. and Gon\ifmmode \mbox{\c{c}}\else \c{c}\fi{}alves, L.F. and Cardman, R. and Raithel, G.},
  journal = {Phys. Rev. Appl.},
  volume = {17},
  issue = {4},
  pages = {044020},
  numpages = {6},
  year = {2022},
  month = {Apr},
  publisher = {American Physical Society},
  doi = {10.1103/PhysRevApplied.17.044020},
  url = {https://link.aps.org/doi/10.1103/PhysRevApplied.17.044020}
}

@article{LP2,
  title = {Closed-loop quantum interferometry for phase-resolved Rydberg-atom field sensing},
  author = {Berweger, Samuel and Artusio-Glimpse, Alexandra B. and Rotunno, Andrew P. and Prajapati, Nikunjkumar and Christesen, Joseph D. and Moore, Kaitlin R. and Simons, Matthew T. and Holloway, Christopher L.},
  journal = {Phys. Rev. Appl.},
  volume = {20},
  issue = {5},
  pages = {054009},
  numpages = {8},
  year = {2023},
  month = {Nov},
  publisher = {American Physical Society},
  doi = {10.1103/PhysRevApplied.20.054009},
  url = {https://link.aps.org/doi/10.1103/PhysRevApplied.20.054009}
}

@article{LP3,
  title = {All-Optical Radio-Frequency Phase Detection for Rydberg Atom Sensors Using Oscillatory Dynamics},
  author = {Schmidt, Matthias and Bohaichuk, Stephanie M. and Venu, Vijin and Wang, Ruoxi and K\"ubler, Harald and Shaffer, James P.},
  journal = {Phys. Rev. Lett.},
  volume = {135},
  issue = {9},
  pages = {093602},
  numpages = {6},
  year = {2025},
  month = {Aug},
  publisher = {American Physical Society},
  doi = {10.1103/23kb-7h7q},
  url = {https://link.aps.org/doi/10.1103/23kb-7h7q}
}

\end{document}